\documentclass{emulateapj}

\newcommand{\actaa}{Acta Astron.}

\newcommand{\dg}{\ensuremath{^\circ}}
\newcommand{\cori}{\ensuremath{I^\prime_0}}

\shorttitle{THE STRUCTURE OF THE LMC STELLAR HALO}
\shortauthors{PEJCHA \& STANEK}

\begin{document}

\title{The Structure of the Large Magellanic Cloud Stellar Halo\\
Derived Using OGLE-III RR Lyr stars}

\author{Ond\v{r}ej Pejcha and K.~Z.~Stanek}
\affil{Department of Astronomy, The Ohio State University, 140 West 18th Avenue, Columbus, OH 43210, USA}
\email{pejcha, kstanek@astronomy.ohio-state.edu}

\begin{abstract}
We use the recently released OGLE-III catalog of 17692 fundamental mode RR~Lyr (RRab) stars in the Large Magellanic Cloud (LMC) to investigate the structure of its stellar halo. We apply conservative cuts in period, amplitude and magnitude to remove blends and other contamination. We use period--luminosity and period--color relations to determine distance and extinction of every star in our final sample of 9393 stars. In order to determine the scatter of our method, we compare the distributions of distances in two regions at the edges of the covered area with a central region. We determine the intrinsic line-of-sight dispersion in the center to be 0.135 mag or 3.21 kpc (FWHM of 0.318 mag or 7.56 kpc), assuming zero depth in one of the edge regions. The conservative cuts we apply reduce the derived depth significantly. Furthermore, we find that the distribution of RRab stars is deformed in the sense that stars on the Eastern side are closer than on the Western side. We model the RRab distribution as a triaxial ellipsoid and determine its axes ratios to be $1\!:\!2.00\!:\!3.50$ with the longest axis inclined by $6\dg$ from the line of sight. Another result of our analysis is an extinction map of the LMC and a map of internal reddening, which we make publicly available.
\end{abstract}

\keywords{galaxies: structure --- Magellanic clouds --- stars: variables: other}

\section{Introduction}

The structure and evolution of the Large Magellanic Cloud (LMC) stellar populations has been thoroughly investigated using many different methods \citep[e.g.][]{marel02,nikolaev04,lah05,cole05}. One of the least studied populations remains the stellar halo, for which the preferred tracer objects are old metal-poor RR~Lyr variable stars \citep{kinman91}. \citet{minniti03} and \citet{borissova04,borissova06} measured radial velocity dispersions of a number of RR~Lyr stars to be $\sim 50\ {\rm km\ s}^{-1}$, compared to e.g.,  $20\ {\rm km\ s}^{-1}$ for carbon stars \citep{marel02}. This suggests that RR~Lyr stars form a kinematically hot halo with a significant vertical spread.

In this study, we use the OGLE-III catalog of the fundamental mode RR~Lyr (RRab) stars \citep{sos09} to investigate the structure of the stellar halo of the LMC. The advantage of RRab stars is that distances to individual stars can be estimated using the period--luminosity relation and the extinction can be determined using the period--color relation.

In Section~\ref{sec:method}, we describe the data set and how distance and extinction are estimated for individual stars. In Section~\ref{sec:3d}, we present the extinction map and the three-dimensional structure of the LMC. In Section~\ref{sec:dis}, we discuss the possible systematics of our method, compare our findings to previous results on the structure of the LMC and discuss the implications of our results.

\section{Data and method}
\label{sec:method}

\begin{figure*}[t]
\plotone{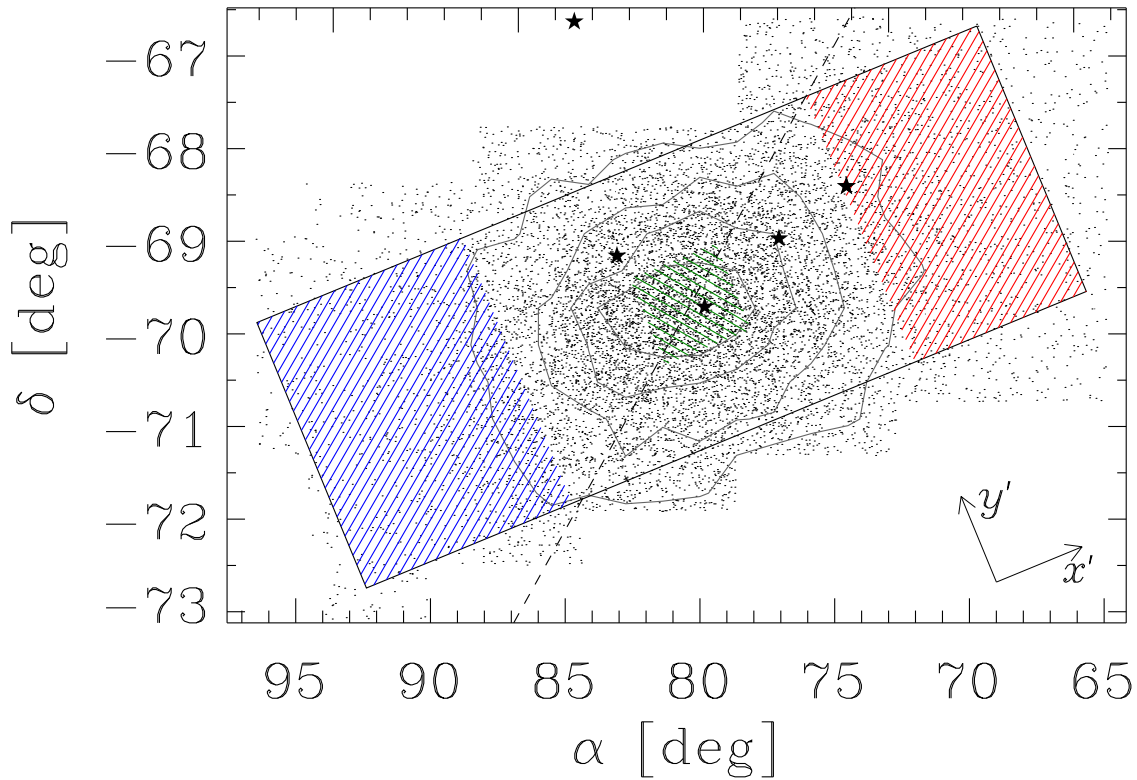}
\caption{Positions of the $11171$ RRab stars satisfying cuts on period, amplitude and magnitude. Density contours marking levels of 250, 500, 750 and 1000 stars per square degree are shown with grey lines. The rectangle containing the $9393$ stars of our final sample is shown in black. Regions that are investigated more closely are marked with blue, green and red parallel lines. The line of nodes measured by \citet{nikolaev04} is shown with a black dashed line going through the centroid position of our sample $(\alpha_0, \delta_0)$. Black filled stars show positions of eclipsing binaries with measured distances; in order of increasing $\alpha$: HV~2274 \citep{ribasetal00,groenewegen01}, OGLE-0510 \citep{pietrzynskietal09}, EROS~1044 \citep{ribasetal02}, HV~982 \citep{fitzpatricketal02} and HV~5936 \citep{fitzpatricketal03}.}
\label{fig:fields}
\end{figure*}

To study the structure of the LMC old stellar population we use $9393$ RRab stars. We start with $17692$ candidate RRab stars selected from the catalog of \citet{sos09}, who also provided the pulsation mode classification based on Fourier coefficients of the light curves. We then exclude stars that were flagged as uncertain, foreground, or belonging to one of the LMC globular clusters. We applied further cuts to reduce the number of blended stars, likely misclassifications and stars with extreme properties. Specifically, we keep only stars with periods $0.45\ {\rm days}\leq P \leq 0.70\ {\rm days}$, amplitudes $0.30\ {\rm mag} \leq \Delta I \leq 0.85\ {\rm mag}$ and $I \geq 18$~mag. Furthermore, we remove the approximately $9\%$ of stars that fall outside the main period--amplitude sequence shown in Figure~3 of \citet{sos09}. The equatorial positions $(\alpha, \delta)$ of the remaining $11171$ stars are plotted in Figure~\ref{fig:fields}. The edges of the OGLE-III coverage are rather irregular with respect to the RR~Lyr distribution. To prevent appearance of artifacts in projected density plots we remove stars that fall outside a rectangle of size $8.4\dg \times 3.1\dg$ and oriented parallel to the long axis of RRab distribution on the sky. This leaves $9393$ stars as our final sample. The rectangle is shown in black in Figure~\ref{fig:fields} and the method for determining the shape of the RRab distribution is discussed in Section~\ref{sec:3d}. The centroid position of our final sample is $(\alpha_0, \delta_0) = (80.35\dg,\ 69.68\dg)$.

In order to estimate distances to individual RRab stars in the LMC we make use of the period--luminosity and period--color relations determined from RRab stars in the galactic globular cluster M3. We take advantage of the fact that the properties of RRab stars in this cluster are similar to the LMC, namely the mean period and the position of ridgelines in the period--amplitude diagram \citep{alcock00,clement08}. Furthermore, the extinction toward M3 is very small, $E(B-V) \approx 0.01$ mag \citep{cacciari05,schlegel98}. For the slope of the period--luminosity relation $\eta$ we use the value determined by \citet{benko06}, namely $\eta =1.453$. They also found that the scatter of their period--luminosity relation is $0.04$ mag. From the dataset of \citet{benko06} we determined the relation between period and intrinsic color $(V-I)_0$ using robust linear least-squares fit. The period--color relation is
\begin{equation}
(V-I)_0 = 0.69 + 0.89\log P
\label{eq:vizero}
\end{equation}
with an rms scatter of $0.062$ mag.

The $I$-band extinction $A_I$ to a particular star is estimated from the difference between the observed color index $(V-I)$ and the intrinsic color using
\begin{equation}
A_I = R_{VI}[(V-I) - (V-I)_0],
\label{eq:ext}
\end{equation}
where $R_{VI} = R_I/(R_V-R_I)$ is the ratio between total $I$-band extinction and reddening of the $(V-I)$ color. If we get $A_I < 0$ from Equation~(\ref{eq:ext}), we manually set $A_I = 0$. Throughout this paper, we set $R_{VI}=1.10$, a slightly lower value than the usual $R_{VI} = 1.55$ \citep[e.g.][]{schlegel98}, to avoid inconsistencies in extinction estimates. We discuss implications of different values of $R_{VI}$ in Section~\ref{sec:syst}. 

Using the information provided we are able to correct the observed intensity-mean $I$ magnitude for extinction $A_I$ and for period--luminosity relation according to 
\begin{equation}
\cori = I - \eta \log P - A_I.
\label{eq:corr}
\end{equation}
The line-of-sight distance $z$ to a particular star is then obtained from $\cori$ assuming that the median distance of our final sample is $50$ kpc. Our method resembles the approach of \citet{nikolaev04} who studied the LMC Cepheids.

\begin{figure*}[t]
\center{
\includegraphics[width=0.7\textwidth]{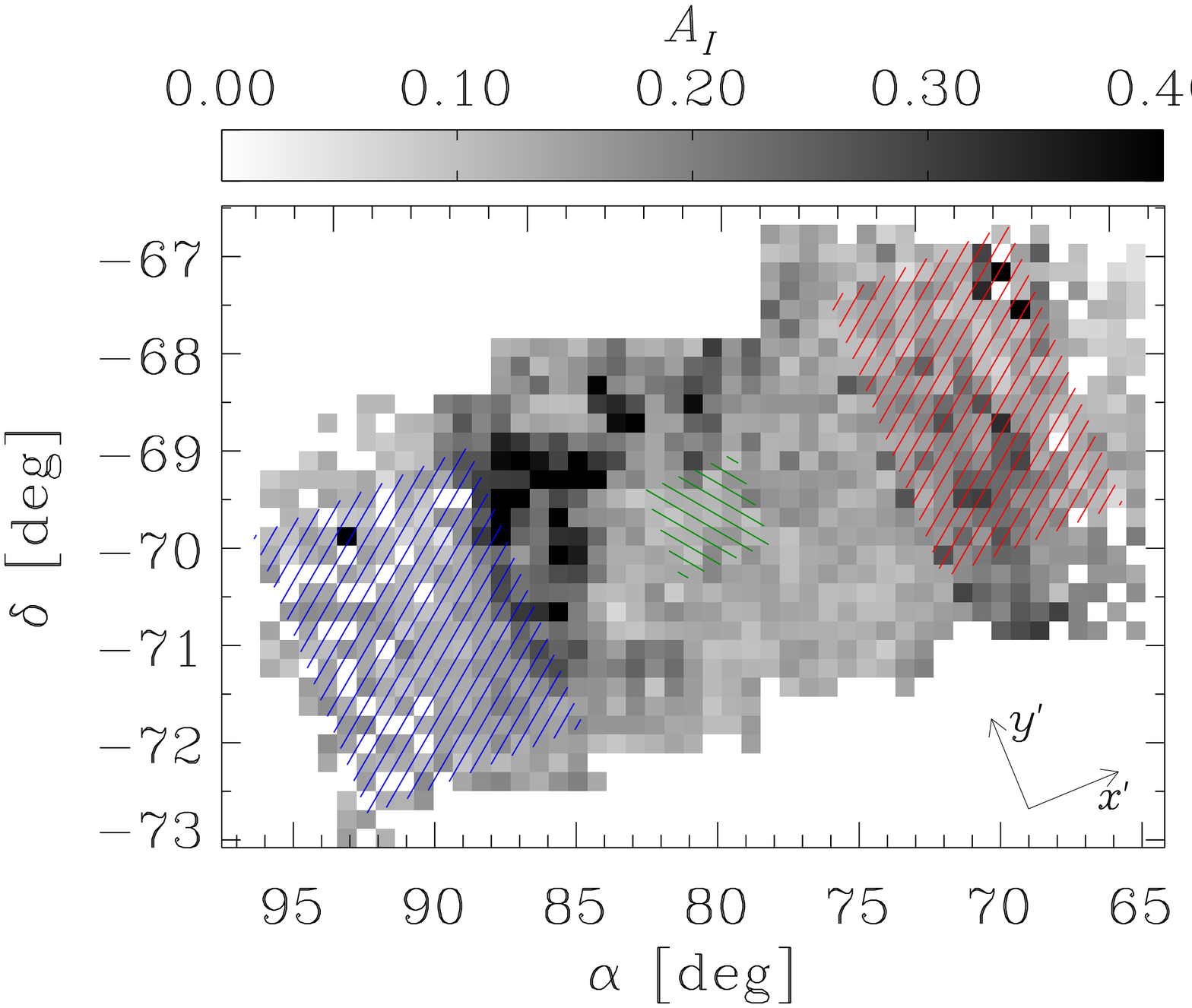}\\
\includegraphics[width=0.7\textwidth]{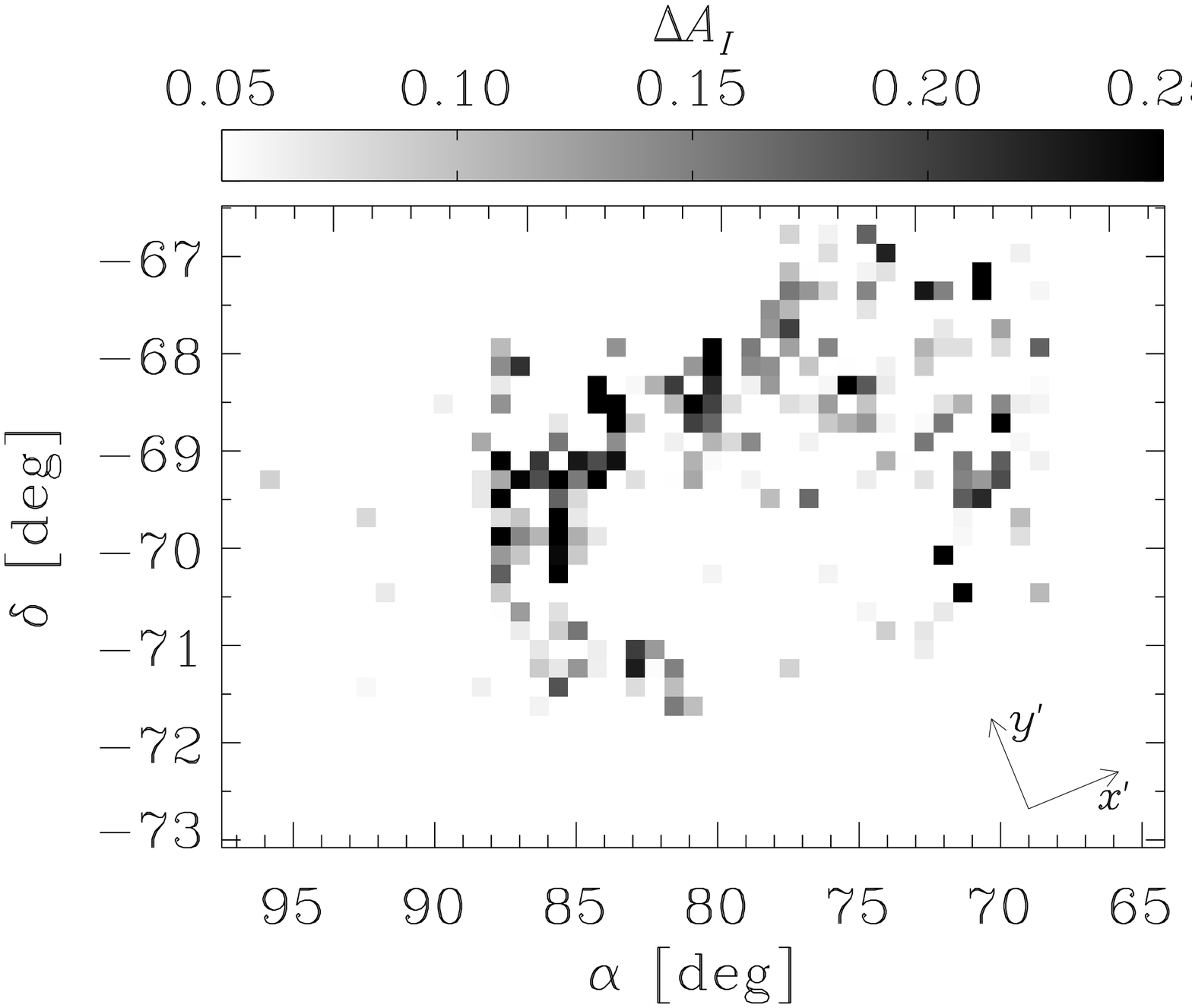}
}
\caption{{\em Upper panel}: Extinction map of the LMC. We calculated the mean value of $A_I$ in pixels of size $0.2\dg \times 0.2\dg$ containing at least two RRab stars. Otherwise, we assign $A_I=0$. The blue, green and red regions from Figure~\ref{fig:fields} are marked here, too. {\em Lower panel}: Extinction variation along the line of sight $\Delta A_I$. We plot only pixels with $\Delta A_I \geq 0.05$ mag. }
\label{fig:extmap}
\end{figure*}

\section{The Three-dimensional Structure of the LMC}
\label{sec:3d}

In this section, we first create an extinction map of the LMC and estimate the internal reddening. Then we focus on the three-dimensional structure of the stellar halo of the LMC as traced by RRab stars.

\subsection{Extinction map and internal reddening}
\label{sec:extmap}

We construct the extinction map\footnote{The extinction map can be obtained at \url{http://www.astronomy.ohio-state.edu/\~{}pejcha/lmc\_extmap} } of the LMC by dividing the plane of the sky into pixels of $0.2\dg \times 0.2\dg$ and calculating mean $A_I$ for every pixel containing at least two stars. The extinction map is plotted in the upper panel of Figure~\ref{fig:extmap}. Comparison with the extinction maps of \citet{zaritsky04} reveals the same basic structure. There is a high-extinction ($A_I \gtrsim 0.4$~mag) region near 30~Doradus extending toward north--west and a north--south filament of increased extinction ($A_I \sim 0.25$~mag) at $\alpha \sim 70\dg$. Our map also shows the low extinction central region seen in the \citet{zaritsky04} maps. 

\begin{figure*}[t]
\plotone{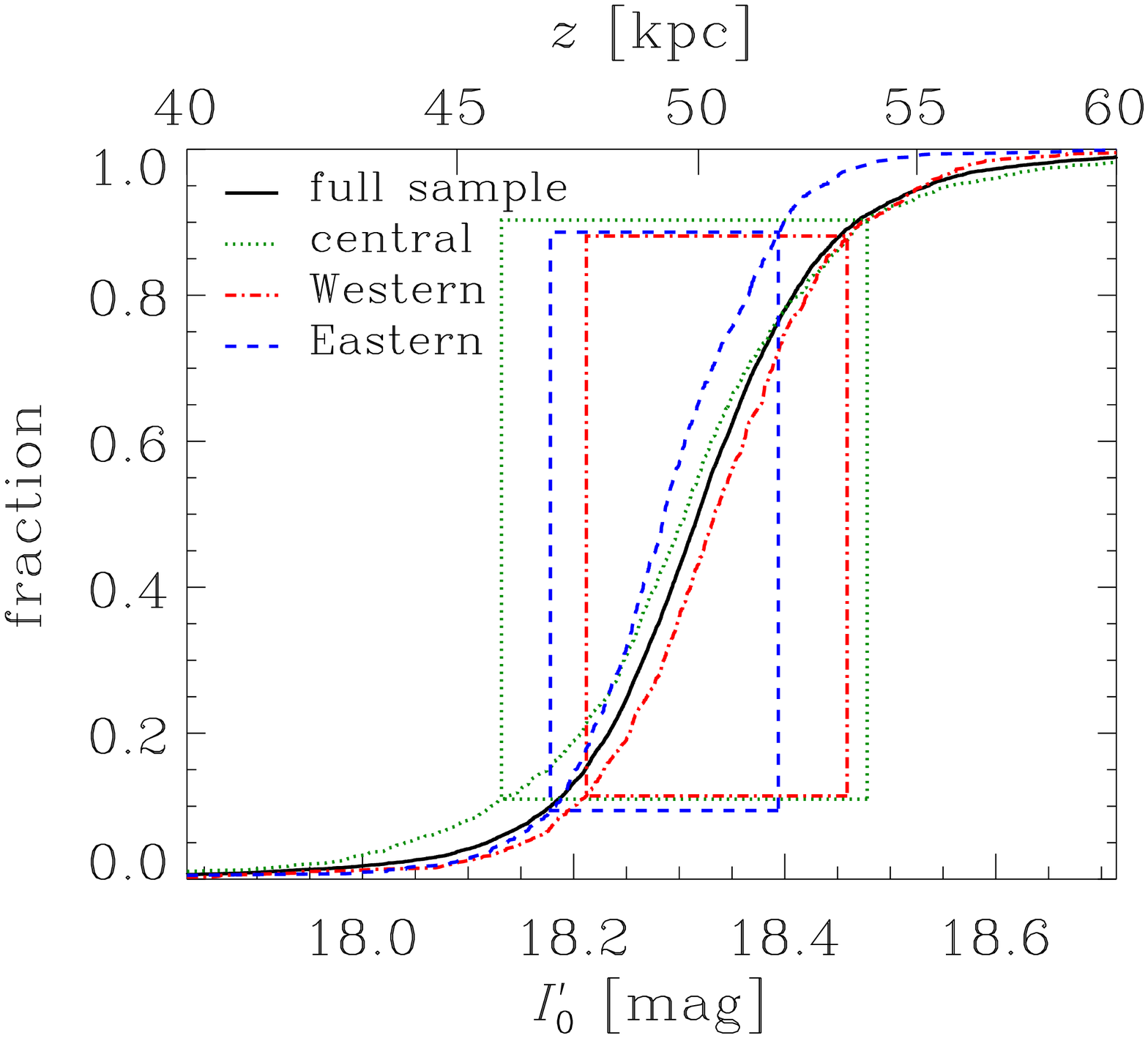}
\caption{Cumulative distributions of extinction-corrected RRab magnitudes $\cori$ for all $9393$ stars (black solid line) and those in the regions marked in Figure~\ref{fig:fields}. The Eastern region is shown with dashed blue line, central region with dotted green line and Western region with dash-dotted red line). The boxes near the center of the plot denote areas of $68\%$ confidence areas around the mean.}
\label{fig:cumul}
\end{figure*}

Equipped with an estimated distance $z$ to every star, we can estimate the line-of-sight structure of the extinction. To this end, we select pixels with at least four stars and split the stars into two equal groups based on their line-of-sight distance. We calculate the mean $A_I$ of each group and then subtract these two values to obtain the difference in extinction $\Delta A_I$ between distant and close stars. The values of $\Delta A_I$ are positive if the more distant stars have higher extinction. The distribution of values of $\Delta A_I$ is skewed to positive values with robust standard deviation of $0.063$ mag and skewness of $1.48$. We plot $\Delta A_I$ as a function of position on the sky in the lower panel of Figure~\ref{fig:extmap}. As expected, the high-extinction regions seen in the upper panel of Figure~\ref{fig:extmap} show up also in the internal reddening plot in the lower panel. A similar map of internal reddening was constructed by \citet{subramanian09} from the spread of colors of red clump stars measured by the Magellanic Cloud Photometric Survey. Their map also shows high internal reddening close to 30~Doradus, but the north--south high-extinction filament at $\alpha \sim 70\dg$ is not clearly visible.

\subsection{Structure of the Stellar Halo}
\label{sec:3dhalo}

In order to determine scatter of our distance estimation method we first select two regions (the blue and red areas in Figure~\ref{fig:fields}) containing about $900$ stars and lying near the edges of the covered area where we expect the intrinsic depth to be lowest. We can then attribute the measured width of the distance distribution to the internal scatter of our distance estimation method. Cumulative distributions of $\cori$ for the selected regions are shown in Figure~\ref{fig:cumul} with dashed blue and dash-dotted red lines, respectively. The widths of the distributions of $\cori$ measured by the standard deviations are $0.108$~mag ($2.5$~kpc at $z=50$ kpc) and $0.123$~mag ($2.9$~kpc) for the Eastern and Western regions, respectively, and are shown in Figure~\ref{fig:cumul} with blue dashed and red dash-dotted rectangles. We assume that the scatter of our distance estimation method is that of the Eastern region, i.e., $0.108$~mag or $2.5$~kpc. We note that the scatter of period--luminosity and period--color relations gives $0.08$~mag ($1.9$~kpc) as the scatter of our distance estimation method. Hence, our estimate of scatter in the distances is conservative.

\begin{figure*}[t]
\plotone{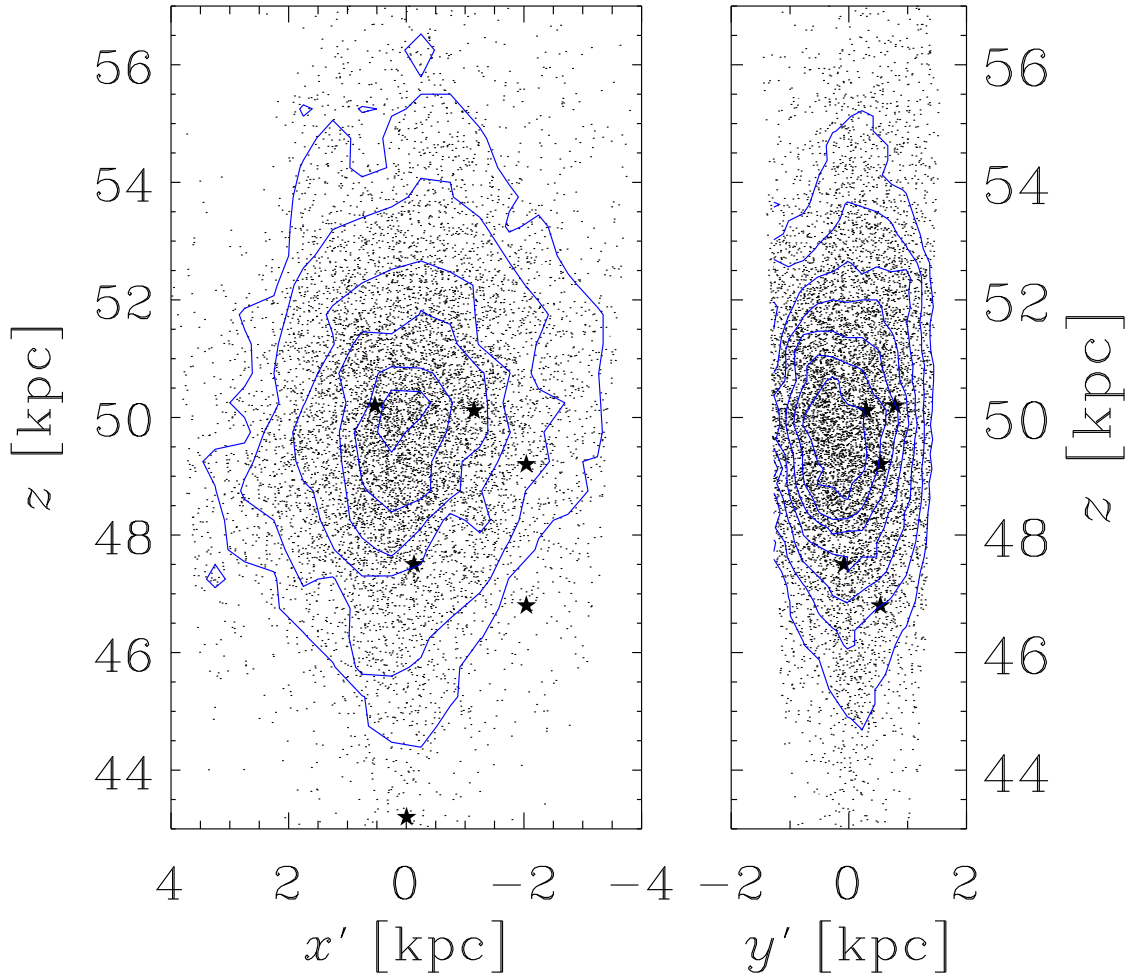}
\caption{Line-of-sight distance $z$ of stars in our final sample as a function of rotated coordinates $x'$ and $y'$. Blue lines mark contours of density $50, 100, 200, 300, 400$ and $500$ kpc$^{-2}$ for the $(x',z)$ plane and $100, 200, 300, 400, 500, 600$ and $700$ kpc$^{-2}$ for the $(y',z)$ plane. Black stars mark the positions of the eclipsing binaries shown in Figure~\ref{fig:fields}; note that we plot two distances for HV~2274 \citep{groenewegen01,ribasetal00} and that HV~5936 falls outside the limits of the right panel.}
\label{fig:proj}
\end{figure*}

We expect the depth of the LMC to be highest close to the center. We plot in Figure~\ref{fig:cumul} cumulative distribution (green dotted line) of $\cori$ of about $1100$ stars in a $1\dg \times 1\dg$ region centered on $(\alpha_0, \delta_0)$. The cumulative distribution is significantly wider than that of the side regions and also of the full sample that is shown with black line. The standard deviation of this subsample is $0.173$ mag ($4.1$~kpc) and is shown in Figure~\ref{fig:cumul} with a green dotted rectangle. Using the scatter of our distance estimation method determined above, we can determine the intrinsic dispersion of depth in the central region to be $0.135$ mag or $3.21$ kpc (FWHM of $0.318$ mag or $7.56$ kpc). As the intrinsic depth in the Eastern region most likely differs from zero, we can consider this value as a lower estimate. For comparison, the standard deviations of the distribution along axes of symmetry in the plane of the sky are $1.52$ and $0.89$ kpc, respectively. This means that the LMC RRab population depth is roughly twice the spatial extent in the plane of the sky. 

From Figure~\ref{fig:cumul}, we also see that the distributions for the Eastern and Western regions are shifted with respect to each other, in the sense that the Eastern region is on average about $0.05$~mag ($1.1$ kpc) closer to us than the Western region. Indeed, stellar populations of the LMC are known to be a part of an disk with a line of nodes at the position angle $\theta = 151.0^\circ \pm 2.4^\circ$ \citep{nikolaev04} and inclined so that the Eastern side is closer to the observer. From the RRab stars, we now see that the halo is also deformed so that the Eastern side is closer. Encouraged by these result, we now examine the full three-dimensional distribution of RRab stars.

The population of RRab stars in the LMC can be approximated by a triaxial ellipsoid. To determine the axis ratios and space orientation of the ellipsoid we first convert equatorial coordinates for every star to physical coordinates $(\alpha, \delta) \rightarrow (x,y)$ with the centroid of our sample corresponding to the coordinate origin, $(\alpha_0,\delta_0) \rightarrow (0,0)$, and keeping $x$ to increase to the east. We then construct the covariance matrix of the coordinates (essentially the inertia tensor) and determine its eigenvalues and eigenvectors. The ratios of the square roots of the eigenvalues are the ratios of the axes and the eigenvectors correspond to their spatial directions \citep[for a similar analysis see e.g.][]{paz06}. We first apply this method only in the plane of the sky described by the $(x,y)$ coordinates and find that the distribution of RRab stars is elongated with an axis ratio of $1.71$ and a major axis position angle of $112.4^\circ$. We use this result to define new rotated coordinates,  $(x,y)\rightarrow (x', y')$. Lines parallel to the longer and shorter sides of the rectangle in Figure~\ref{fig:fields} correspond to $x'$ and $y'$ axes, respectively. The results for the distribution in the plane of the sky are robust as they depend only on the precisely measured positions.

Now we repeat the same procedure for the $(x,y,z)$ coordinates, however, here the line-of-sight distance $z$ is smeared by the internal scatter of our method. To correct for this, we subtract the value of internal scatter determined above from the appropriate element of the covariance matrix. We obtain the axis ratios of  $1\!:\!2.00\!:\!3.50$, where the longest axis is inclined by $6\dg$ from the line of sight and the position angle of the projection of the ellipsoid on the plane of the sky is $113.4^\circ$.

The finite sky coverage of our data can bias the determination of axis ratios and directions. To see how serious this problem is we remove stars lying outside the $250$ stars per square degree contour shown in Figure~\ref{fig:fields} and repeat the procedure. We get the axis ratios of $1\!:\!1.36\!:\!3.53$ with the longest axis inclined by $3\dg$ from the line of sight. The decrease in the relative length of the second longest axis is caused by removing relatively more stars along $x'$ than along $y'$.

\begin{figure}[t]
\plotone{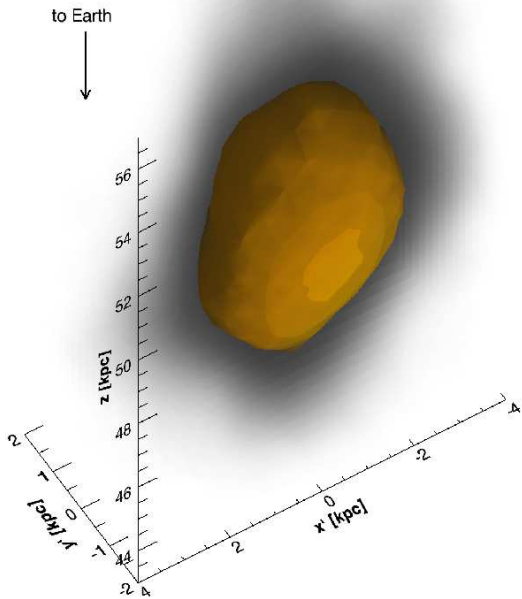}
\caption{Three-dimensional distribution of LMC RRab stars in $(x', y', z)$ coordinates. The full distribution is displayed in semi-transparent gray and an isodensity surface of $56$ kpc$^{-3}$ is shown in dark yellow. The visualization was constructed from $0.5\times 0.5 \times 0.5$ kpc$^{-3}$ cubic voxels smoothed with a boxcar width of two elements.}
\label{fig:loaf}
\end{figure}

Figure~\ref{fig:proj} shows the spatial distribution of RR~Lyr stars in the $(x',z)$ and $(y', z)$ planes with overplotted density contours. Looking on the left panel we confirm our result that the Eastern part (with $x' >0$) is on average closer to us than the Western part. In Figure~\ref{fig:loaf} we give a more intuitive visualization of the spatial distribution of RRab stars in the LMC using a three-dimensional plot of an isodensity surface. 

\section{Discussion and conclusion}
\label{sec:dis}

In this section, we first discuss possible systematics that might be present in our analysis, then we compare our results to previous studies and conclude with implications for the structure of the LMC and future work.

\newpage
\subsection{Possible systematics}
\label{sec:syst}

There are a number of possible systematic errors that could lower the significance of our results. First of all, the assumption underlying all of our reasoning is that the population of RRab stars is homogeneous throughout the LMC. While we have no a priori reason to assume this, there are several checks for systematic errors. For example, comparing the period distribution of stars in our three regions, we find a mean period difference of about $0.003$ days, or about $0.01$ mag in the period--luminosity relation. Another check is that the distributions of light curve Fourier coefficients show no differences between these regions above the level of $3\%$ difference in the mean. In short, we do not see any evidence of different RR~Lyr populations across the LMC. 

In our extinction correction of Equation~(\ref{eq:ext}), we set $R_{VI}$ to a value lower than the fiducial $R_{VI} = 1.55$. We repeated our analysis with several values in the range $1.00 \leq R_{VI} \leq 1.55$ and we found that the distance distribution widths tend to increase with increasing $|R_{VI}-1.10|$. For example, for $R_{VI} = 1.55$ we found standard deviations of $0.118$, $0.118$ and $0.191$ mag for our Eastern, Western, and central regions, respectively.  Furthermore, for values of $R_{VI} \gtrsim 1.2$ we find inconsistencies, in the sense that stars closer to us have a higher average extinction than the more distant stars. Or in other words, the mean $\Delta A_I$ was found to be about $-0.016$~mag compared to $0.028$~mag for $R_{VI} = 1.10$. Lower values of ratio of total to selective extinction to the LMC were found also by \citet{misselt99}.

The cuts we applied on period, amplitude and magnitude tend to reduce the outliers and consequently the derived dispersion in depth. For example, applying no cuts yields standard deviation of $0.35$ mag for the central region and $0.20$ mag for the side regions; also the central region is on average significantly brighter than the side regions for all reasonable values of $R_{VI}$. Making our cuts even more stringent does not yield significantly smaller dispersions (or reduce the difference between the central and side regions). From this analysis we conclude that our cuts efficiently remove contaminated stars and that contamination of our sample is insignificant.

We did not apply any cuts on color in order not to bias the derived shape in case there were large-scale extinction variations. However, applying an additional color cut of $0.3 \leq (V-I) \leq 0.9$ removes highly reddened stars, but does not change our results. Specifically, the intrinsic line-of-sight standard deviation in the central region would be $0.123$~mag ($2.91$~kpc) and the axis ratio would be $1\!:\!1.99\!:\!3.14$ with the longest axis inclined by $9\dg$ from the line of sight. The application of the color cut changes our results on the intrinsic depth and axis ratios and directions only slightly.

\subsection{Comparison with Other Work}
\label{sec:comp}

\citet{minniti03} and \cite{borissova04,borissova06} obtained radial velocities of $137$ RR~Lyr stars and found a very high velocity dispersion of about $50\ {\rm km\ s}^{-1}$ suggesting that the stars form a kinematically hot stellar halo. Our results on the depth dispersion are comparable with the results of \citet{clementini03}, who analyzed RR~Lyr stars in two regions close to the bar and found a depth dispersion between $3.3$ kpc and $3.8$ kpc, especially if we realize that the estimate of intrinsic scatter of our method is based on outer parts of the halo which probably have substantial depth too. \citet{subramaniam06} used a smaller sample of RR~Lyr stars from OGLE-II \citep{sos03} to derive the scale-height of $3.0 \pm 0.9$ kpc, which translates to dispersion of about $6.0 \pm 1.8$ kpc. This is higher than our result, the difference stems most likely from the fact that we estimated extinction correction on a star-by-star basis and we applied stringent cuts to remove contaminations. 

\citet{nikolaev04} found a number of Cepheids lying significantly ($>7$\,kpc) both in front and behind the main disk of the LMC. Similarly, the distance determinations to eclipsing binaries in the LMC show scatter larger than the uncertainties \citep[e.g.][]{groenewegen01,fitzpatricketal03,pietrzynskietal09}. Both types of objects are considerably younger and more metal rich than RR~Lyr stars, but it is possible that the LMC halo contains a fraction of younger objects. A population of objects in a non-spherical halo elongated along the line of sight could help to explain the number of observed microlensing events towards the LMC \citep{alcock00b,evans00,alves04,calchi06,wyrzykowski09}. As the microlensing cross-section depends basically linearly on the lens--source distance, doubling this distance with constant halo mass yields roughly twice cross-section of the self lensing due to halo of the LMC. 

\subsection{Conclusion and future work}

We have investigated the structure of the LMC halo using fundamental mode RR~Lyr stars. We found that the halo line-of-sight depth is variable, in the sense that the depth is higher in the central areas than at the edges. Furthermore, the eastern side of the halo is in general closer to us than the western side, mimicking in part the LMC disk. We were also able to provide extinction maps of the LMC and resolve the internal extinction.

While there is a significant increase of covered area between OGLE-II and OGLE-III catalogs, a survey of RR~Lyr stars far from the axis of symmetry of their distribution on the sky would help to better characterize their spatial distribution. It is possible that far from the center the distribution of RR~Lyr stars might exhibit an overdensity similarly to Cepheids, that are organized in a ring with an off-center bar \citep{nikolaev04}. 

Furthermore, \citet{borissova06} found little variation in the velocity dispersion as a function of distance from the center of rotation, but their sample has only two stars in the farthest $[2.0; 2.5]$ kpc bin. The OGLE-III catalogue \citep{sos09} gives stars as far as $5$ kpc from the center and it would be interesting to see whether the difference of the line-of-sight depth between the central and side regions appears also in radial velocity measurements.

\acknowledgments We are grateful to the OGLE team for making their data publicly available. We are grateful to C.~S.~Kochanek for detailed and useful comments and careful reading of the manuscript. We thank S.~Dong and I.~Soszy\'nski for discussions and careful reading of the manuscript, and to the anonymous referee for suggestions.

\clearpage
\end{document}